\documentclass{article}
\usepackage[pdftex]{graphicx}
\usepackage{amsmath}

\begin{document}

\title{Quantum incompressibility of a falling Rydberg atom, and a
gravitationally-induced charge separation effect in superconducting systems}
\author{Raymond Y. Chiao \\
Schools of Natural Sciences and of Engineering\\
University of California\\
P. O. Box 2039\\
Merced, CA 95344\\
E-mail: rchiao@ucmerced.edu}
\date{March 7a, 2010}
\maketitle

\begin{abstract}
Freely falling point-like objects converge towards the center of the Earth.
Hence the gravitational field of the Earth is inhomogeneous, and possesses a
tidal component. The free fall of an extended quantum object such as a
hydrogen atom prepared in a high principal-quantum-number stretch state,
i.e., a circular Rydberg atom, is predicted to fall more slowly that a
classical point-like object, when both objects are dropped from the same
height from above the Earth. This indicates that, apart from
\textquotedblleft quantum jumps,\textquotedblright\ the atom exhibits a kind
of \textquotedblleft quantum incompressibility\textquotedblright\ during
free fall in inhomogeneous, tidal gravitational fields like those of the
Earth.

A superconducting ring-like system with a persistent current circulating
around it behaves like the circular Rydberg atom during free fall. Like the
electronic wavefunction of the freely falling atom, the Cooper-pair
wavefunction is \textquotedblleft quantum incompressible.\textquotedblright\
The ions of the ionic lattice of the superconductor, however, are not
\textquotedblleft quantum incompressible,\textquotedblright\ since they do
not possess a globally coherent quantum phase. The resulting difference
during free fall in the response of the \emph{nonlocalizable} Cooper pairs
of electrons and the \emph{localizable} ions to inhomogeneous gravitational
fields is predicted to lead to a charge separation effect, which in turn
leads to a large repulsive Coulomb force that opposes the convergence caused
by the tidal, attractive gravitational force on the superconducting system.

A \textquotedblleft Cavendish-like\textquotedblright\ experiment is proposed
for observing the charge separation effect induced by inhomogeneous
gravitational fields in a superconducting circuit. This experiment would
demonstrate the existence of a novel coupling between gravity and
electricity via macroscopically coherent quantum matter.
\end{abstract}

\pagebreak

\section{\protect\bigskip Introduction}

Experiments at the frontiers of quantum mechanics and gravity are rare. I
would like to explore in this essay in honor of Danny Greenberger and Helmut
Rauch, situations which could lead to such experiments. The key is to
understand the phenomenon of \textquotedblleft quantum
incompressibility\textquotedblright\ of macroscopically coherent quantum
matter in the presence of inhomogeneous, tidal gravitational fields, such as
the Earth's. See Figure 1.

\begin{figure}[ptb]
\begin{center}
\includegraphics[width=4in]{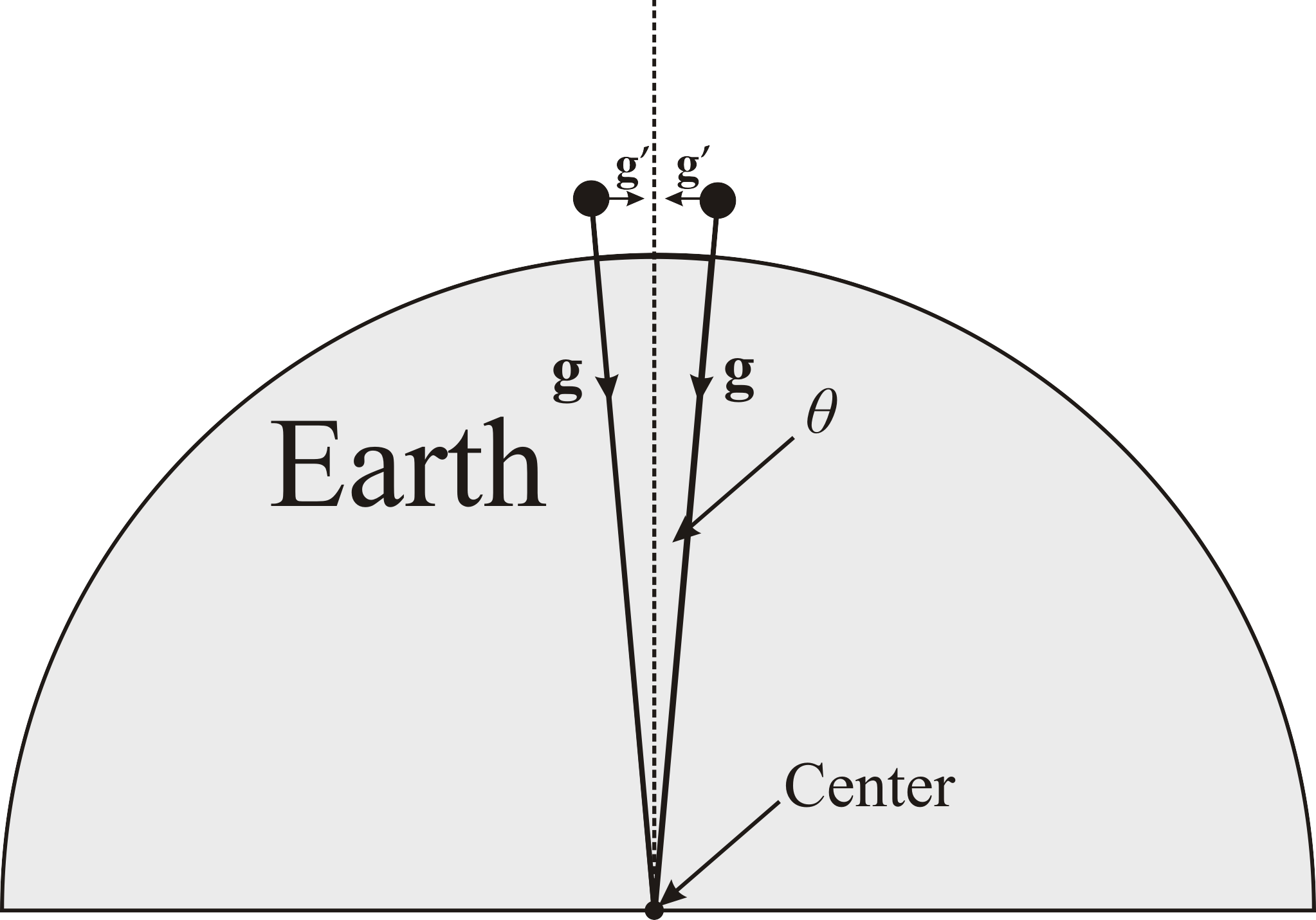}
\end{center}
\caption{Two nearby, freely falling, point-like objects dropped from the
same height above the Earth's surface follow converging trajectories that
are inclined at a slight angle $\protect\theta $ with respect to the
vertical plumb line equidistant between them. According to a distant
inertial observer, the radial convergence of these objects' trajectories
towards the center of the Earth causes them to undergo small horizontal
components of acceleration $\mathbf{g}^{\prime }$ of the radial acceleration
$\mathbf{g}$. These components are equivalent to a tidal gravitational force
that, in a Newtonian picture, causes the two objects to converge toward one
another.}
\end{figure}

As an example of \textquotedblleft quantum
incompressibility\textquotedblright\ during free fall of an extended quantum
object,\ let us first consider the single electron of a circular Rydberg
atom \cite{Hulet-and-Kleppner} (ignoring electron spin), which is prepared
in the state
\begin{equation}
\left\vert n,\text{ }l=n-1,\text{ }m=n-1\right\rangle \text{ ,}
\label{ring-like state}
\end{equation}
where $n$ is the principal quantum number, which is large, i.e., $n\gg 1$,
and $l=n-1$ is the maximum possible orbital angular momentum quantum number
for a given $n$, and $m=l=n-1$ is the maximum possible azimuthal quantum
number for a given $l$, i.e., the \textquotedblleft
stretch\textquotedblright\ state. The $z$ axis has been chosen to be the
local vertical axis located at the center of mass of the atom. Then the
wavefunction of this electron in polar coordinates $(r,\theta ,\phi )$ of
the hydrogenic atom in this state is given by \cite{Haroche-and-Raimond}
\begin{equation}
\Psi _{n,n-1,n-1}(r,\theta ,\phi )=N_{n,n-1,n-1}(r\sin \theta e^{i\phi
})^{n-1}\exp \left( -\frac{r}{na_{0}}\right) \text{ ,}  \label{ring-like wf}
\end{equation}
where $N_{n,n-1,n-1}$ is a normalization constant. The probability density
associated with this wavefunction has the form of a strongly peaked
distribution which lies on the horizontal $(x,y)$ plane, in the shape of a
ring of radius
\begin{equation}
a_{n}=n^{2}\frac{\hbar ^{2}}{me^{2}}=n^{2}a_{0}\text{ ,}
\label{size of Rydberg atom}
\end{equation}
where $a_{0}$ is the Bohr radius. Thus one recovers the Bohr model of the
hydrogen atom in the correspondence-principle limit of large $n$. This
ring-like probability distribution is illustrated in Figure 2. \qquad
\begin{figure}[ptb]
\begin{center}
\includegraphics[width=4in]{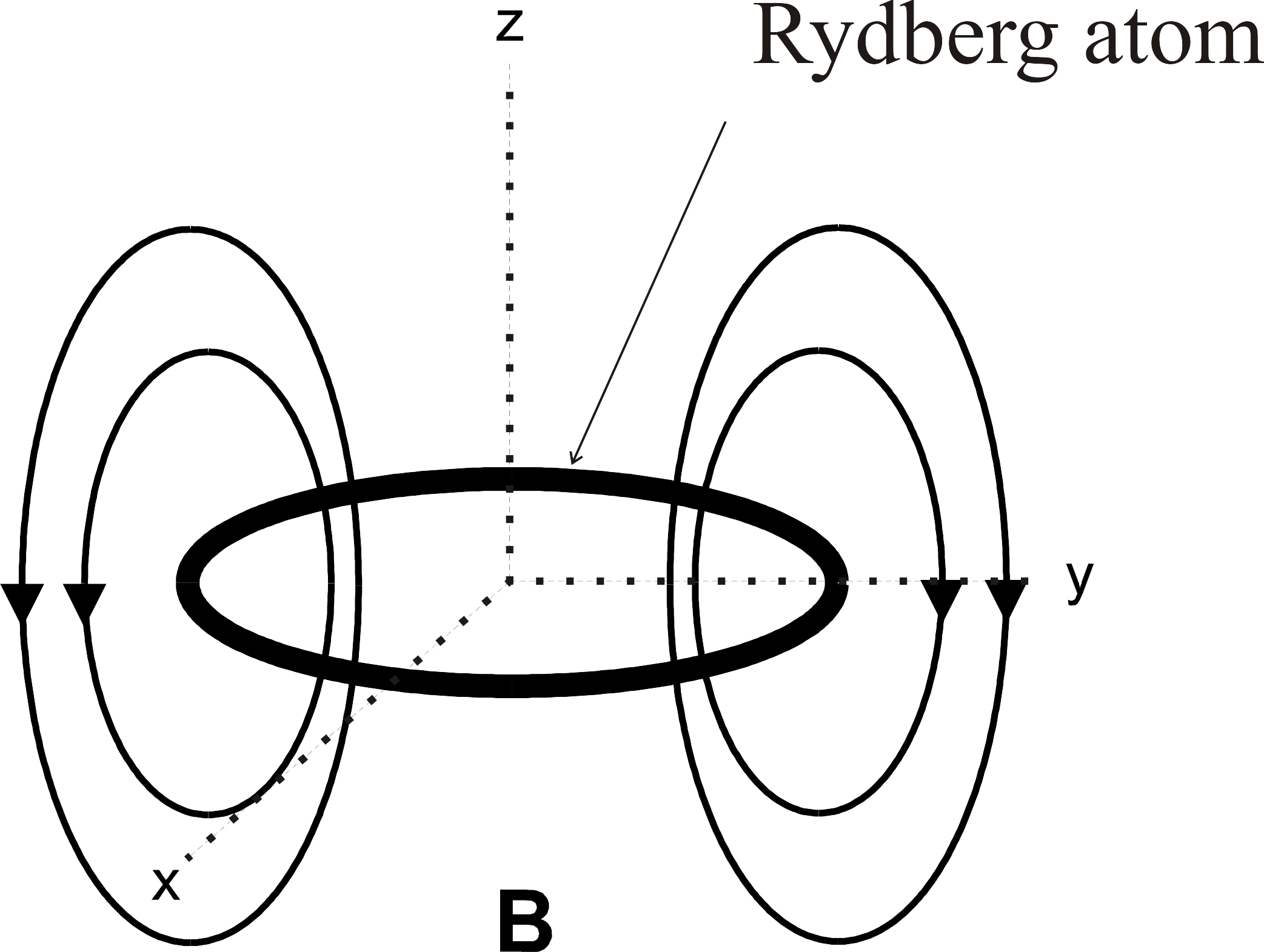}
\end{center}
\caption{A circular Rydberg atom in the state $\left\vert n,\text{ }l=n-1,%
\text{ }m=n-1\right\rangle $ has a strongly peaked, ring-like probability
distribution, i.e., an \textquotedblleft electron cloud,\textquotedblright\
indicated by the heavy black loop. Currents in this state lead to a magnetic
field $\mathbf{B}$, indicated by the directional loops. The $z$ axis is the
local vertical axis.}
\end{figure}

The question I would like to address here is this: How does the size of this
atom change with time as it undergoes free fall in Earth's inhomogeneous,
tidal gravitational field?

\section{\protect\bigskip An analogy}

The magnetic moment of the Rydberg atom in the state (\ref{ring-like wf}) is
quantized, and is given by
\begin{equation}
\mu _{n}=n\frac{e\hbar }{2m}=n\mu _{B}\text{ ,}
\end{equation}
where $\mu _{B}$ is the Bohr magneton and $n$ is an integer.

The electron current density in the ring-like structure of a circular
Rydberg atom in Figure 2 is similar to that of a persistent supercurrent of
Cooper pairs in a superconducting ring with a quantized flux given by
\begin{equation}
\Phi _{n}=n\Phi _{0}=n\frac{h}{2e}\text{ ,}
\end{equation}
where $\Phi _{0}$ is the flux quantum and $n$ is an integer. The quantum
incompressibility of the ring-like structure of a circular Rydberg atom, and
the quantum incompressibility of the Cooper pairs of electrons in a
superconducting ring, both arise from the same quantum mechanical principle,
namely, the single-valuedness of the wavefunction after one round trip
around the ring, which follows from the condition
\begin{equation}
\oint\limits_{\text{ring}}\nabla \varphi \cdot d\mathbf{l}=\Delta \varphi
=2\pi m\text{ ,}
\end{equation}
where $\varphi $ is the phase of the wavefunction, and $m$ is an integer
corresponding to the state under consideration. Another necessary condition
for quantum incompressibility is the existence of a substantial energy gap
separating the $m$th state from adjacent states of the system.

The analogy between the Rydberg atom and the superconducting ring is not a
perfect one, since the selection rules for allowed transitions between
adjacent states will be different in the two cases. The transitions $%
n\rightarrow n-1$ and $n\rightarrow n+1$ are electric-dipole allowed for the
Rydberg atom, whereas the transitions $n\rightarrow n-1$ and $n\rightarrow
n+1$ between adjacent flux-trapping states of the superconducting ring are
highly forbidden. This is because a macroscopic number of identical Cooper
pairs of electrons must all simultaneously jump from a state with $n\hbar $
units to a state with $\left( n-1\right) \hbar $ units or with $\left(
n+1\right) \hbar $ units of angular momentum per electron pair. Hence the
persistent current of a superconducting ring is highly metastable, and does
not change with time, unless a macroscopic \textquotedblleft quantum
jump\textquotedblright\ occurs.

If the characteristic frequency of an external perturbation, such as that of
the tidal gravitational fields acting on the system during free fall in
Earth's gravity, is much less than the smallest energy gap of an allowed
transition divided by Planck's constant, then the system cannot make a
transition (i.e., a \textquotedblleft quantum jump\textquotedblright ) out
of its initial state. Thus it must stay rigidly in its initial state. (For a
Rydberg atom with $n\simeq 100$, this transition frequency lies in the
gigahertz range, so that this assumption is well satisfied.) The size of the
circular Rydberg atom and the size of the persistent currents of the
superconducting ring will therefore remain constant in time during
perturbations arising from Earth's tidal fields during free fall, apart from
a sequence of possible \textquotedblleft quantum jumps\textquotedblright\ in
a \textquotedblleft quantum staircase.\textquotedblright

\section{The quantum incompressibility of the Rydberg atom}

Let us show that quantum incompressibility is predicted to occur in a
circular Rydberg atom, starting from DeWitt's minimal coupling rule. The
DeWitt Hamiltonian for a freely falling hydrogenic atom, such as a circular
Rydberg atom in presence of weak electromagnetic and gravitational fields,
is given in SI units by
\begin{equation}
H=\frac{1}{2m}\left( \mathbf{p}-e\mathbf{A}-m\mathbf{h}\right) ^{2}+\frac{
e^{2}}{4\pi \varepsilon _{0}r}\text{ ,}
\label{DeWitt Hamiltonian for Rydberg atom}
\end{equation}
where $\mathbf{A}$ is the electromagnetic vector potential, and $\mathbf{h}$
\ is DeWitt's gravitational vector potential \cite{DeWitt}.

Let us first briefly review the simpler case when a DC magnetic field is
turned on without any accompanying gravitational field, i.e., when $\mathbf{%
A }\neq \mathbf{0}$ and $\mathbf{h}=\mathbf{0}$. The interaction Hamiltonian
for the $\mathbf{A\cdot A}$ term (the \textquotedblleft Landau diamagnetism
term\textquotedblright ) is given by\ \cite{Landau&Lifshitz}\cite{A.p}
\begin{equation}
H_{\mathbf{A\cdot A}}=\frac{e^{2}}{2m}\mathbf{A\cdot A}\text{ .}
\end{equation}
In the symmetric gauge, where $\mathbf{A}=\frac{1}{2}\mathbf{B}\times
\mathbf{r=-}\frac{1}{2}B(y\mathbf{e}_{x}-x\mathbf{e}_{y})$, for $\mathbf{B}%
=B \mathbf{e}_{z}$, and where $\mathbf{e}_{x},\mathbf{e}_{y},$ and $\mathbf{e%
} _{z}$ are the unit vectors along the $x$, $y$, and $z$ axes, respectively,
this yields
\begin{equation}
H_{\mathbf{A\cdot A}}=\frac{e^{2}B^{2}}{8m}(x^{2}+y^{2})\text{ .}
\end{equation}%
The energy shift in first-order perturbation theory resulting from the
presence of the $\mathbf{A}$ field is given by%
\begin{equation}
\Delta E_{\mathbf{A\cdot A}}=\frac{e^{2}B^{2}}{8m}\left\langle \Psi
_{nlm}\left\vert x^{2}+y^{2}\right\vert \Psi _{nlm}\right\rangle \text{ .}
\label{energy A}
\end{equation}
Recalling that the wavefunction for the circular Rydberg state is given by 
(\ref{ring-like wf}), the expectation value in (\ref{energy A}) becomes
\begin{equation}
\left\langle \Psi _{n,n-1,n-1}\left\vert x^{2}+y^{2}\right\vert \Psi
_{n,n-1,n-1}\right\rangle \approx (n^{2}a_{0})^{2}=a_{n}^{2}\text{ }
\label{mean-square-size}
\end{equation}
for large values of the principal quantum number $n$, where $n>>1$. It
follows that the first-order energy shift of the atom in the presence of a
magnetic field is
\begin{equation}
\Delta E_{\mathbf{A\cdot A}}\approx \frac{e^{2}a_{n}^{2}}{8m}\text{ }B^{2}
\text{.}  \label{A.A_energy_shift}
\end{equation}
This result implies that, in first-order perturbation theory, the size of
the atom does not change in the presence of the applied DC magnetic field,
in the sense that the root-mean-square transverse size of the atom, which is
given by
\begin{equation}
\left. a_{n}\right\vert _{\text{rms}}=\sqrt{\left\langle \Psi
_{n,n-1,n-1}\left\vert x^{2}+y^{2}\right\vert \Psi _{n,n-1,n-1}\right\rangle
}=a_{n}
\end{equation}
does not change with time during the application of the DC magnetic field.
Moreover, all the moments of the atomic probability distribution do not
change, since the wavefunction $\Psi _{n,n-1,n-1}$ remains unaltered in
first-order perturbation theory in the presence of a weak applied field.
Furthermore, this is still true for applied magnetic fields which vary
sufficiently slowly in time, so that no transitions (i.e., \textquotedblleft
quantum jumps\textquotedblright ) can occur out of the initial state of the
system $\Psi _{n,n-1,n-1}$. The concept of the \textquotedblleft quantum
incompressibility\textquotedblright\ of a Rydberg atom thus is a valid
concept during the application of sufficiently weak, and sufficiently slowly
varying, magnetic fields.

The energy shift given by (\ref{A.A_energy_shift}) causes the atom to become
a \emph{low-field seeker} in inhomogeneous magnetic fields through the
relationship
\begin{equation}
\left( \mathbf{F}_{_{\mathbf{A\cdot A}}}\right) _{n}=-\nabla \left( \Delta
E_{\mathbf{A\cdot A}}\right) _{n}\approx -\frac{e^{2}a_{n}^{2}}{8m} {\nabla \left( B^{2}\right)}
\end{equation}%
where $\left( \mathbf{F}_{_{\mathbf{A\cdot A}}}\right) _{n}$ is the force on
the atom in the ring-like state (\ref{ring-like wf}) in the presence of an
inhomogeneous magnetic field.

Next, let us consider the more interesting case of when weak tidal
gravitational fields are present without any accompanying electromagnetic
fields, i.e., when $\mathbf{h}\neq \mathbf{0}$ and $\mathbf{A}=\mathbf{0}$.
As before, the atom is initially prepared in the state given by (\ref%
{ring-like wf}) before it is released into free fall in the Earth's
inhomogeneous gravitational field. The $z$ axis, which goes through the
center of mass of the atom, is chosen to be the local vertical axis of the
Earth's field. The horizontal tidal gravitational fields of the Earth
experienced during free fall by the atom, as observed in the coordinate
system of a distant inertial observer, where the $(x,y)$ plane is the local
horizontal plane, will be given by
\begin{equation}
\mathbf{h}(x,y,t)=\mathbf{v}(x,y,t)=\mathbf{g}^{\prime }t=\frac{gt}{R_{E}}(
\mathbf{e}_{x}x+\mathbf{e}_{y}y)\text{ ,}  \label{tidal h field}
\end{equation}
where $\mathbf{v}(x,y,t)$ is the velocity of a freely falling, point-like
test particle located at $(x,y)$ and observed at time $t$ by the distant
inertial observer \cite{Prague}, $\mathbf{g}^{\prime }$ is the horizontal
component of Earth's gravitational acceleration arising from the radial
convergence of free-fall trajectories towards the center of the Earth as
seen by this observer (see Figure 1), $R_{E}$ is the radius of the Earth,
and $\mathbf{e}_{x}$ and $\mathbf{e}_{y}$ are respectively the unit vectors
pointing along the $x$ and the $y$ axes, in this observer's coordinate
system. In (\ref{tidal h field}) we have assumed that the horizontal
excursions of the electron in $x$ and $y$ are very small compared to the
Earth's radius. The interaction Hamiltonian for the $\mathbf{\ h}\cdot
\mathbf{h}$ term in (\ref{DeWitt Hamiltonian for Rydberg atom}) is given by
\begin{equation}
H_{\mathbf{h}\cdot \mathbf{h}}=\frac{m}{2}\mathbf{h}\cdot \mathbf{h}=\frac{
mg^{2}t^{2}}{2R_{E}^{2}}(x^{2}+y^{2})\text{ .}
\end{equation}
Therefore, the shift in energy of the atom in the circular Rydberg state,
due to the Earth's tidal fields given by (\ref{tidal h field}), is given in
first-order perturbation theory by
\begin{equation}
\Delta E_{\mathbf{h}\cdot \mathbf{h}}=\frac{mg^{2}t^{2}}{2R_{E}^{2}}
\left\langle \Psi _{n,n-1,n-1}\right\vert x^{2}+y^{2}\left\vert \Psi
_{n,n-1,n-1}\right\rangle \approx \frac{ma_{n}^{2}}{2R_{E}^{2}}g^{2}t^{2}
\label{energy shift
for h}
\end{equation}
for large values of $n$, where $n>>1$. Once again, since the expectation
value in (\ref{energy shift for h}) is the mean-square transverse size of
the atom, this implies that the size of the atom does not change during free
fall, according to first-order perturbation theory. In other words, the atom
is \textquotedblleft quantum incompressible\textquotedblright\ in the
presence of the inhomogeneous, tidal fields of the Earth, just like in the
case of the atom in the presence of an applied DC magnetic field, as long as
transitions (i.e., \textquotedblleft quantum jumps\textquotedblright ) out
of the initial quantum state $\Psi _{n,n-1,n-1}$\ cannot occur. This
conclusion is valid assuming that the characteristic frequency of the
applied tidal fields is much less than the gap frequency (i.e., the energy
gap divided by Planck's constant, which is typically on the order of
gigahertz for $n\sim 100$) corresponding to a quantum transition from the $n$
th state to the nearest adjacent allowed states, and assuming that the tidal
gravitational field of the Earth is sufficiently weak.

In the gravitational case, just as in the magnetic case, the energy-level
shift caused by the tidal perturbations arising from the Earth's
inhomogeneous gravitational field, leads to a force on the atom. This force
causes the atom to become a \emph{low-field seeker} in the inhomogeneous
gravitational\emph{\ }field of the Earth through the relationship
\begin{equation}
\left( \mathbf{F}_{_{\mathbf{h\cdot h}}}\right) _{n}=-\nabla \left( \Delta
E_{\mathbf{h\cdot h}}\right) _{n}\approx -\frac{1}{2}ma_{n}^{2}t^{2}\text{ }
\nabla \left( \frac{g^{2}}{R_{E}^{2}}\right) \text{ .}  \label{F_h.h}
\end{equation}
Thus a hydrogen atom in a circular Rydberg state, which is an \emph{extended}
quantum object, will fall slightly more slowly than a point-like classical
test particle which is simultaneously released into free fall along with the
atom near the center of mass of the atom in Earth's inhomogeneous field.

The gravitational, Landau-like energy shifts of the atom given by (\ref%
{energy shift for h}) are much too small to be measured directly in
Earth-bound experiments using current technology, but in principle they can
be measured spectroscopically by monitoring the frequencies of transitions
between adjacent Rydberg states, for example, in a satellite laboratory
which is in a highly elliptical orbit around the Earth. It is therefore a
genuine physical effect.

One might be tempted, as a result of the force given by (\ref{F_h.h}), to
question the universal applicability of the equivalence principle, i.e., the
\emph{universality} of free fall. But it must be kept in mind that the
equivalence principle applies strictly only to \emph{point-like} objects,
inside of which tidal effects can be neglected. This is manifestly not the
case for the \emph{extended} quantum systems being considered here.

\section{A superconducting circuit consisting of two cubes joined coherently
by two parallel wires}

The analogy between the Rydberg atom and superconducting ring suggests a
simple experiment to test the idea of \textquotedblleft quantum
incompressibility\textquotedblright\ during free fall, which can be
performed in an ordinary laboratory. Consider a horizontal system consisting
of two superconducting cubes joined by two parallel superconducting wires to
form a superconducting circuit, which is \emph{topologically} equivalent to the 
circular Rydberg atom. See Figure 3.

\begin{figure}[ptb]
\begin{center}
\includegraphics[width=4in]{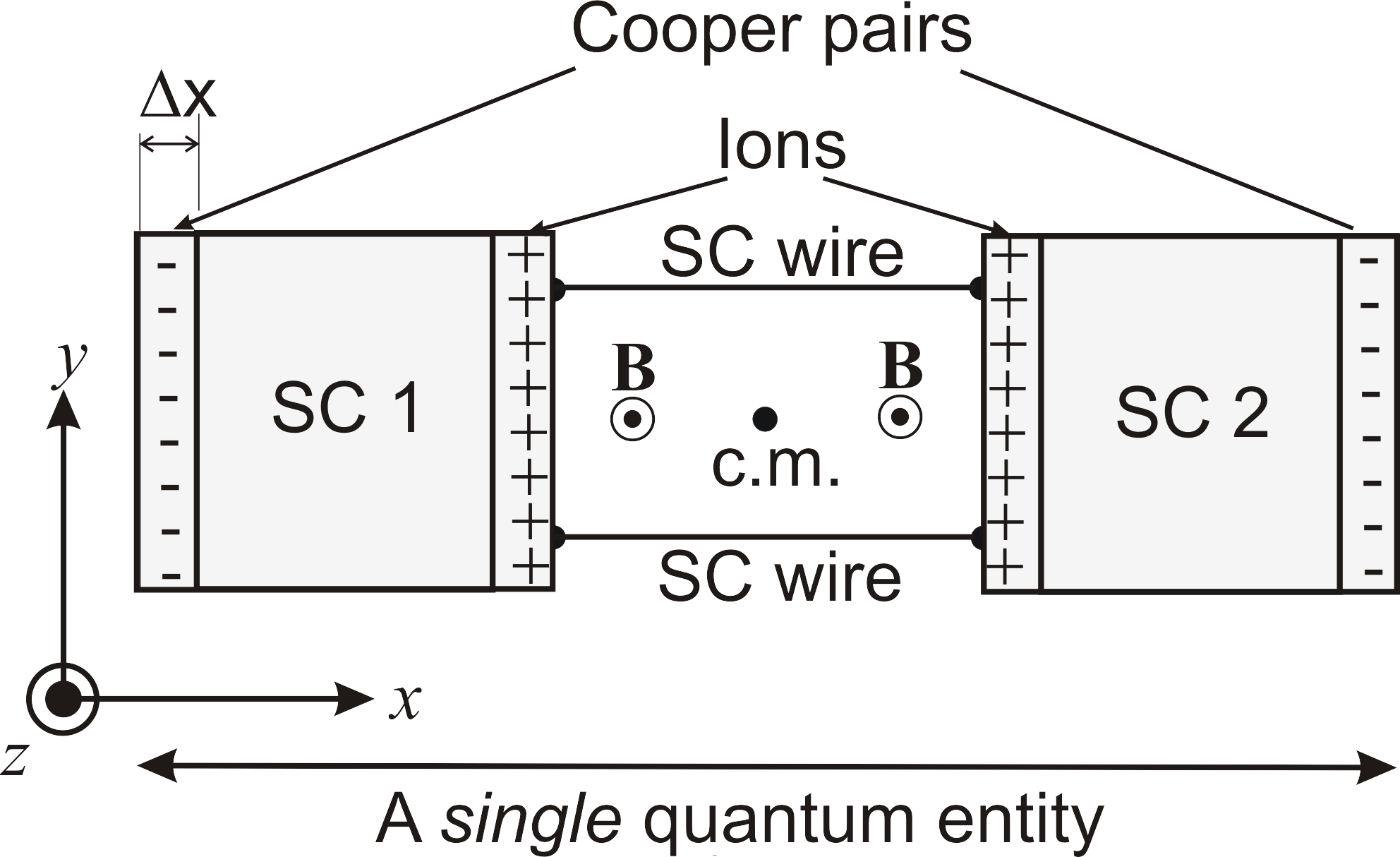}
\end{center}
\caption{Two superconducting cubes, SC 1 and SC 2, which are undergoing free
fall in Earth's inhomogeneous gravitational field, are connected by means of
two thin superconducting wires, which establishes \emph{quantum coherence}
throughout the system, and makes it a single quantum entity with a center of
mass (\textquotedblleft c.m.\textquotedblright ) located in the middle.\ A
persistent current through the wires traps a $\mathbf{B}$ field inside this
superconducting circuit, much like in the circular Rydberg atom. All
dimensions of the cubes and the length of the wire are given by the same
distance $L$. The $z$ axis denotes the local vertical axis passing through
\textquotedblleft c.m.\textquotedblright .}
\end{figure}

When a coherent quantum connection between the two cubes is not present
(due, say, to the effect of heating coils wrapped around the midsections of
both wires which drive them normal by heating them above their transition
temperature, so that the coherent quantum connection between the cubes is
thereby destroyed), the centers of masses of the two spatially separated
cubes, which will have decohered with respect to each other, will follow the
converging free-fall trajectories shown in Figure 1, which are inclined at a
slight angle $\theta $ with respect to the vertical plumb line passing
through the midpoint \textquotedblleft c.m.\textquotedblright , with
\begin{equation}
\theta \approx \frac{L}{R_{E}}\text{ ,}
\end{equation}
where $L$ is separation of the two cubes, which is also their dimensions
(thus chosen for simplicity), and $R_{E}$ is the radius of the Earth. It
should be noted that the \emph{decoherence}, and therefore the \emph{\
spatially separability}, of entangled states arising from perturbations due
to the environment \cite{Zurek}, is a necessary precondition for the
applicability of the equivalence principle \cite{Prague}, so that here the
universality of free fall can be applied to the free-fall trajectories of
the \emph{disconnected} superconducting cubes \cite{geodesics}.

When a coherent connection \emph{is }present between the two cubes, they
will become a single, macroscopic quantum object like the freely-falling
Rydberg atom. The Cooper pairs of electrons of the system will then remain
motionless with respect to the midpoint \textquotedblleft
c.m.\textquotedblright , since their macroscopic wavefunction corresponds to
a zero-momentum eigenstate relative to this \textquotedblleft
c.m.\textquotedblright , and therefore, by the uncertainty principle, the
electrons are completely nonlocalizable within the entire, coherently
connected two-cube system. The Cooper pairs of electrons, like the electron
in Rydberg atom, will then exhibit quantum incompressibility\ during free
fall. This follows from the fact that the wavefunction remains unaltered by
the perturbation, and that therefore all moments of the probability
distribution, and hence the mean-squared size of the coherent electrons of
the \emph{entire} superconducting system, remains unchanged in response to
the tidal gravitational fields of the Earth, according to first-order
perturbation theory.

However, the ions, which have undergone decoherence due to the environment
\cite{Zurek}\cite{Prague}, are completely \emph{localizable}, and therefore,
by the equivalence principle, will want to follow the free-fall trajectories
that converge onto the center of the Earth shown in Figure 1. By contrast,
the Cooper pairs of electrons will remain \emph{coherent} during free fall,
since they are protected from decoherence by the BCS energy gap \cite{Prague}%
, and will therefore remain completely \emph{nonlocalizable}, since they
will remain in a zero-momentum eigenstate. This difference in the motion of
the ions and of the Cooper pairs of electrons will then lead to the
charge-separation effect indicated in Figure 3, in which the ions will be
extruded through the innermost faces of the cubes, because of the
convergence of their radial trajectories that point towards the center of
the Earth, and in which the Cooper pairs of electrons, which resist this
convergence, will be extruded through the outermost faces of the cubes.

Therefore the Cooper pairs of electrons in the zero-momentum eigenstate,
which remain at rest rigidly with respect to the \emph{global}
\textquotedblleft c.m.\textquotedblright\ of the entire, coherent two-cube
system, will be displaced away from the ions by a distance $\Delta x$ on
left face of the left cube, and also on the right face of the right cube.
The resulting charge configuration can be approximated by a ball-and-stick
model of two charged dumbbells shown in Figure 4.

\begin{figure}[ptb]
\begin{center}
\includegraphics[width=4in]{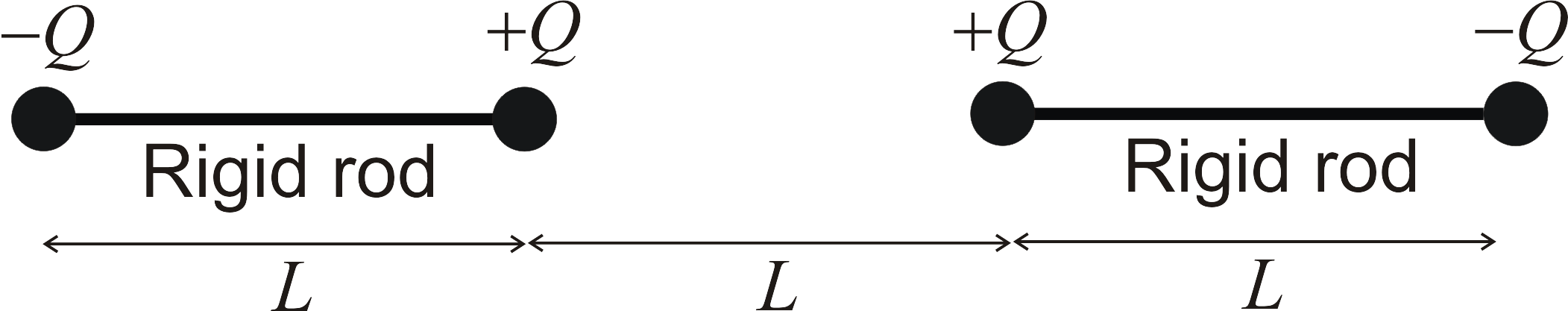}
\end{center}
\caption{Ball with charges $-Q$ and $+Q$ are attached to rigid rods with
lengths $L$ to form two dumbbells, which model the configuration of charges
in Figure 3. The two innermost charges, both of which are $+Q$, are
separated by a distance $L$. These two innermost charges dominate the
Coulomb force between the two dumbbells, so that the net force is a
repulsive one.}
\end{figure}

On the one hand, the net Coulomb force between the two dumbbells in Figure 4
is given by
\begin{equation}
F_{\text{Coulomb}}=\alpha \frac{Q^{2}}{4\pi \varepsilon _{0}L^{2}}\text{ ,}
\label{alpha}
\end{equation}
where $\alpha $ is a pure number on the order of unity (this follows from
dimensional considerations, since $L$ is the only distance scale in the
problem).

On the other hand, the tidal gravitational force between the cubes in Figure
3 is given by
\begin{equation}
F_{\text{Tidal}}=Mg^{\prime }\text{ ,}
\end{equation}
where $M$ is the mass of the cube (which is mainly due to the ions), and
\begin{equation}
g^{\prime }=g\sin \theta \approx g\tan \theta \approx g\theta \approx
gL/R_{E}
\end{equation}
is the horizontal component of the acceleration due to Earth's gravity
acting on the centers of the cubes, which is directed towards the midpoint
\textquotedblleft c.m.\textquotedblright\ of the two cubes. Thus, in
equilibrium,
\begin{equation}
F_{\text{Coulomb}}=F_{\text{Tidal}}\text{ .}
\end{equation}

The voltage difference between the two ends of a given dumbbell (which is a
model of the voltage difference between the opposite faces of a given cube)
is given by
\begin{equation}
V=\beta \frac{Q}{4\pi \varepsilon _{0}L}\text{ ,}  \label{beta}
\end{equation}
where $\beta $ is another pure number of the order of unity (again, this
follows from dimensional considerations, since $L$ is only distance scale in
the problem). Substituting the squared quantity $Q^{2}/L^{2}$ obtained from
( \ref{beta}) into (\ref{alpha}), one gets
\begin{eqnarray}
\alpha \frac{Q^{2}}{4\pi \varepsilon _{0}L^{2}} &=&\alpha \frac{(4\pi
\varepsilon _{0})^{2}V^{2}}{(4\pi \varepsilon _{0})\beta ^{2}}=4\pi
\varepsilon _{0}\frac{\alpha }{\beta ^{2}}V^{2}  \notag \\
&=&Mg^{\prime }\approx \rho L^{3}g\theta \approx \rho g\frac{L^{4}}{R_{E}}
\text{ .}
\end{eqnarray}
Solving for the voltage difference $V$, one obtains
\begin{equation}
V\approx \left( \frac{\beta ^{2}}{\alpha }\frac{\rho gL^{4}}{4\pi
\varepsilon _{0}R_{E}}\right) ^{1/2}=\frac{\left\vert \beta \right\vert }{
\sqrt{\alpha }}V_{\text{F-F}}\text{ ,}
\end{equation}
where the characteristic free-fall voltage scale $V_{\text{F-F}}$ for
characteristic experimental parameters ($L=1$ cm, $\rho =10^{4}$ kg/m$^{3}$)
is given by
\begin{equation}
V_{\text{F-F}}=\left( \frac{\rho gL^{4}}{4\pi \varepsilon _{0}R_{E}}\right)
^{1/2}\sim 1\text{ Volt ,}
\end{equation}
which is experimentally interesting, given the reasonably large capacitances
of the two-cube system. (For the geometry of the dumbbells indicated in
Figure 4, the numerical values of $\alpha =11/18$ and $\beta =-2/3$ are, as
indicated earlier, on the order of unity.)

\section{The \textquotedblleft Cavendish-like\textquotedblright\ experiment}

The order-of-magnitude estimate given above indicates that experiments are
feasible. A team consisting of my new colleague at the University of
California at Merced, Prof. Michael Scheibner, my graduate students, Steve
Minter, Luis Martinez, and Bong-Soo Kang, an undergraduate student, Phil
Jensen, my colleague at Boston University, Prof. Kirk Wegter-McNelly, and I,
are presently performing a \textquotedblleft
Cavendish-like\textquotedblright\ experiment, in which we are producing a
slowly time-varying, inhomogeneous, tidal gravitational field by means of
two piles of lead bricks placed diametrically opposite each other on a
slowly rotating, circular platform, as the sources of the field. The two
piles of bricks, which weigh approximately a ton, will orbit slowly and
symmetrically around a superconducting circuit similar to the one shown in
Figure 3, which is mounted inside a stationary dilution refrigerator that is
suspended above the center of the rotating platform.

We expect to be able to see (using synchronous detection) the charge
separation induced by these gravitational fields in a superconducting
circuit, which consists of two well-separated superconducting bodies, both
of which are suspended by means of pairs of superconducting wires inside the
same refrigerator, so that the two bodies form the superconducting plumb
bobs of two pendula. These two bodies are then \emph{coherently} connected
to each other by means of a pair of parallel superconducting wires, as
indicated in Figure 3, to form a single superconducting circuit, i.e., a
single quantum entity. The charge separation effect can then be measured
inductively by means of a sensitive electrometer (we should be able to see
the charge induced by the extruded Cooper pairs, which should be on the
order of picocoulombs, with a high signal-to-noise ratio). If we should
observe a nonzero charge-separation signal in this experiment, then this
observation would establish the existence of a novel coupling between
gravity and electricity mediated by means of macroscopic quantum matter.

Normally, i.e., for plumb bobs made out of non-superconducting materials, 
the gravitational fields due to the ton of bricks should cause
small angular deflections on the order of nanoradians of the two pendula.
These small deflections should occur relative to the local vertical axis
which is located at the midpoint in between them (see Figure 1, in which the
two freely-falling objects are replaced by the two plumb bobs of the two
pendula). We would normally expect to see such deflections if these two
pendula consisted of normal, classical matter, or if they consist of two
superconducting plumb bobs which have had their superconducting connection
between them destroyed due to decoherence. Such deflections could be
measured with high signal-to-noise ratios using laser interferometry. If we
were to monitor both the deflections of the pendula and the charge
separation effect in the same experiment, there would be four logical
possibilities as to the possible outcomes:

(I) \indent \space \space Charge-separation? YES. \indent Deflection? NO.

(II) \indent \space Charge-separation? NO. \indent \space Deflection? YES.

(III) \indent Charge-separation? YES. \indent Deflection? YES.

(IV) \indent Charge-separation? NO. \indent \space Deflection? NO.

Based on the arguments presented above, we would expect (I) to be the
outcome, if the Cooper pairs were to be able to drag the ions of the lattice
into co-motion with these superconducting electrons during free fall. In the
tug-of-war between the uncertainty principle and the equivalence principle,
the uncertainty principle wins in (I). By contrast, if there is nothing
special about this superconducting system over any other material system,
i.e., if the universality of free fall were to apply to the Cooper pairs
inside the superconducting circuit so that they would undergo free fall
along with the ions, and that therefore the superconducting system would remain
electrically neutral and unpolarized during free fall, then we would expect
(II) to be the outcome. The equivalence principle wins in (II). If, however,
there does exist a charge-separation effect, but the ions of the lattice
were to drag the Cooper pairs into co-motion with the ionic lattice during
free fall, then we would expect (III) to be the outcome. Finally, there
exists the remote possibility of outcome (IV), which would indicate that
Newtonian gravity would somehow have failed to produce any deflection at all
of the pendula in the presence of the ton of bricks.
Results from this \textquotedblleft Cavendish-like\textquotedblright\
experiment will be presented elsewhere.

My heartiest birthday congratulations to Danny and Helmut!
\pagebreak


\begin{thebibliography}{9}
\bibitem{Hulet-and-Kleppner} R. Hulet and D. Kleppner, Phys. Rev. Lett.
\textbf{51}, 1430 (1983).

\bibitem{Haroche-and-Raimond} S. Haroche and J.-M. Raimond, \emph{Exploring
the Quantum} (Oxford University Press,  2006).

\bibitem{DeWitt} B.S. DeWitt, Phys. Rev. Lett. \textbf{16}, 1092 (1966).

\bibitem{Landau&Lifshitz} L.D. Landau and E.M. Lifshitz, \emph{Quantum
Mechanics} (Butterworth-Heinemann, Oxford,  2003).

\bibitem{A.p} In addition to the $\mathbf{A\cdot A}$ and the $\mathbf{h\cdot
h}$ terms, there are also $\mathbf{A\cdot p}$ and $\mathbf{h\cdot p}$ terms
in the perturbation Hamiltonian. However, these latter terms lead to energy
shifts  which vanish in the limit that $n\rightarrow \infty $ in comparison
with the former. For details, see R.Y. Chiao,  S.J. Minter, and K.
Wegter-McNelly (in preparation).

\bibitem{Zurek} W.H. Zurek, Rev. Modern Phys. \textbf{75}, 715 (2003).

\bibitem{Prague} S.J. Minter, K. Wegter-McNelly, and R. Y. Chiao, Physica E
\textbf{42}, 234 (2010).

\bibitem{geodesics} The general relativistic concept of a \textquotedblleft
geodesic\textquotedblright\ is fundamentally  that of a \textquotedblleft
classical trajectory.\textquotedblright\ As Bohr has taught us, however, the
very  concept of a \textquotedblleft classical trajectory\textquotedblright\
loses all physical meaning under  circumstances in which the uncertainty
principle destroys all such classical trajectories. This is the case here
for  the free-fall trajectories of Cooper pairs inside the superconducting
system, since these electron pairs will remain  in a \emph{zero-momentum
eigenstate} during free fall, when they are initially prepared in the BCS
ground state at  the moment of release into free fall.
\end{thebibliography}
\end{document}